\documentclass{article}
\usepackage{amssymb}

\usepackage{graphicx}
\usepackage{amsmath}

\input{tcilatex}

\begin{document}

{\Huge On the Security of the Cha-Ko-Lee-Han-Cheon Braid Group Public-key
Cryptosystem}\bigskip

Milton M. Chowdhury\medskip

\underline{1. Abstract}\medskip

We show that a number of cryptographic protocols using non-commutative
semigroups including the Cha-Ko-Lee-Han-Cheon braid group cryptosystem have
security based on the MSCSP. We give two algorithms to solve the DP using
the MSCSP.\medskip

\underline{2. Introduction}\medskip

At the CRYPTO 2000 conference the seminal KLCHKP
(Ko-Lee-Cheon-Han-Kang-Park)\ braid group public-key cryptosystem was
published see \cite{Ko et al}. An updated version of the KLCHKP cryptosystem
which is the CKLHC (Cha-Ko-Lee-Han-Cheon) braid group cryptosystem was
introduced at ASIACRYPT 2001 conference \cite{Cha et al} the claim of the
authors was the updated cryptosystem is based on the DH-DP (Diffie-Hellman
Decomposition Problem). We show that the KLCHKP and CKLHC cryptosystems are
based on the MSCSP and it has been assumed for several years the security of
these cryptosystems are based on the DH-CP and DH-DP respectively, we also
show the related cryptosystems may be based on the MSCSP and hence give a
new way to break the KLCHKP and CKLHC cryptosystems and the related
cryptosystems for some parameters. It has been shown there is a linear
algebraic attack on the KLCHKP and CKLHC cryptosystems but our attack is
more practical.\medskip

\underline{3. Hard Problems in Non-Abelian Groups}\medskip

Definition-The MSCSP (multiple simultaneous conjugacy search problem) \cite
{Ko} is find elements $g\in G$ such that $y_{i}=gx_{i}g^{-1}$, given the
publicly known information: $G$ is a group, $x_{i},y_{i}\in G$ with $%
x_{i},y_{i}=ax_{i}a^{-1}$, $1\leq i\leq u,$ with the secret element $a\in G.$

Definition-The CSP \cite{Ko} can be defined as the MSCSP with $u=1$.

Notation-We refer an example of the MSCSP as $%
((x_{1},x_{2},...,x_{u}),(y_{1},y_{2},...,y_{u}))$ with solution $g.$\bigskip

The \textbf{DP} (Decomposition Problem) \cite{Shpilrain et al} is defined as
follows.\newline
\textit{Public Information}: $G$ is a semigroup, $A$ is a subset of $G$. $%
x,y\in G$ with $y=axb$.\newline
\textit{Secret information}: $a,b\in A$.\newline
\textit{Objective}: find elements $f,g\in A$ such that $fxg=y$.\medskip 
\newline
The definition of the $DP$ above generalises the definition of a less
general version of the $DP$ given in \cite{Cheon et al}, \cite{Ko} and \cite
{Shpilrain2 et al}. The less general version only differs from the above
definition of $DP$ because $G$ is a group and $A$ is a subgroup. In our
notation in all of this paper we omit the binary operation $\ast $ when
writing products so for example $f\ast x\ast g$ is understood to mean $fxg$.
We require that $\ast $ is efficiently computable.\medskip

The \textbf{CSP} (Conjugacy Search Problem) \cite{AAG}, \cite{Ko} is defined
as follows.\newline
\textit{Public Information}: $G$ is a group. $x,y\in G$ with $y=f^{-1}xf$.%
\newline
\textit{Secret Information}: $f\in G$.\newline
\textit{Objective}: find an element $g\in G$ such that $g^{-1}xg=y$.

Notation-We refer an example of the CSP as $(x,y)$ with solution $g.\medskip 
$

The \textbf{DH-DP} (Diffie-Hellman Decomposition Problem) \cite{Cheon et al}%
, \cite{Ko} is defined as follows.\newline
\textit{Public Information}: $G$ is a group. $A,B$ are subgroups of $G$ with 
$[A,B]=1$. $x,y_{a},y_{b}\in G$ with $y_{a}=axb,$ $y_{b}=cxd$.\newline
\textit{Secret Information}: $a,b\in A$, $c,d\in B$.\newline
\textit{Objective}: find the element $cy_{a}d(=ay_{b}b=acxbd).$\medskip 
\newline
The \textbf{DH-CP} (Diffie-Hellman Conjugacy Problem) is the specialisation
of the DH-DP \cite{Cheon et al} with $a=b^{-1}$ and $c=d^{-1}$.\medskip 
\newline
We now re-define the DP and DH-DP above as used in our key agreement
protocol given in \cite{Chowdhury}. In the rest of this paper below the DP
and DH-DP will mean their re-definitions.\medskip \newline
The re-definition of the \textbf{DP} is as follows.\newline
\textit{Public Information}: $G$ is a semigroup. $A$, $B$ are subsets of $G$%
. $x,y\in G$ with $y=axb$.\newline
\textit{Secret Information}: $a\in A$, $b\in B$.\newline
\textit{Objective}: find elements $f\in A$, $g\in B$ such that $fxg=y$%
.\medskip \newline
The re-definition of the \textbf{DH-DP} is as follows.\newline
\textit{Public information}: $G$ is a semigroup. $A,B,C,D$ are subsets of $G$%
. $x,y_{a},y_{b}\in G$ with $y_{a}=axb$, $y_{b}=cxd$.\newline
\textit{Secret Information}: $a\in A$, $b\in B$, $c\in C$, $d\in D$.$%
[A,C]=1,[B,D]=1$\newline
\textit{Objective}: find the element $cy_{a}d$ $(=ay_{b}b=acxbd)$.\medskip

The EDL problem is to decide if the discrete logarithm of two elements in an
abelian group are the same \cite{Thomas}. The \textbf{EDL} type problem is
as follows \cite{Thomas}.\newline
\textit{Public information}: $G$ is a group. $a,b,y_{a},y_{b}\in G$ with $%
y_{a}=uav$, $y_{b}=wbx$.\newline
\textit{Secret Information}: $u,v,w,$ $x\in G$.\newline
\textit{Objective}: Decide if $F_{a}(y_{a})\cap F_{b}(y_{b})\neq \varnothing 
$. Where $F_{\beta }(\alpha )=\{(a,b)\in B_{n}\times B_{n}$ :$\alpha =a\beta
b\}.\medskip $

We redefine the EDL problem more generally as follows.

\textit{Public information}: $G$ is a Semigroup. $A,B,C,D$ are subsets of $%
G. $ $a,b,y_{a},y_{b}\in G$ with $y_{a}=uav$, $y_{b}=wbx$.\newline
\textit{Secret Information}: $u\in A$, $v\in B$, $w\in C$, $x\in D$.\newline
\textit{Objective}: Decide if $F_{a}(y_{a})\cap F_{b}(y_{b})\neq \varnothing 
$. Where $F_{\beta }(\alpha )=\{(a,b)\in B_{n}\times B_{n}$ :$\alpha =a\beta
b\}.$\medskip

\underline{4. Key Agreement Protocol Using Non-Commutative Semigroups}%
\medskip

In \cite{Chowdhury} we introduced a key agreement protocol and a variant of
it which we briefly describe below.

\begin{itemize}
\item  Phase 0. Initial setup\newline
i) $G$ is chosen and is publicly known.\newline
A first method to select the parameters is to select publicly known subsets
or privately known subsets $L_{A}$, $L_{B}$, $R_{A}$, $R_{B}$ and $Z$ of $G$
are chosen for which either property a) below is true or property b) below
is true. Let $z\in Z$ with $z$ the publicly known element which is the value
of $x$ in the definition of the DH-DP used in the example of the DH-DP in
our new authentication scheme.\newline
Following \cite{Shpilrain2 et al} let $g\in G$ for $G$ a group, $C_{G}(g)$
is the centraliser of $g$, we sketch the modifications to the authentication
scheme (and these apply to the key agreement protocol described below) to
give two further methods to select the subgroups as follows. Publicly known
subsets or privately known $L_{A}$, $L_{B}$, $R_{A}$, $R_{B}$ and $Z$ of $G$
are chosen for which either property a) below is true or property b) below
for the second and third methods below.
\end{itemize}

The second method to select the subgroups is $A$ chooses $(a_{1},a_{2})\in
G\times G$ and publishes the subgroups as a set of generators of the
centralisers $L_{B},R_{B},L_{B}\subseteq C_{G}(a_{1}),R_{B}\subseteq
C_{G}(a_{2}),L_{B}=\{\alpha _{1},...,\alpha _{k}\}$ etc. $B$ chooses $%
(b_{1},b_{2})\in L_{B}\times R_{B}$, and hence can compute $x$ below etc.
Following \cite{Shpilrain2 et al} there is no explicit indication of where
to select $a_{1}$ and/or $a_{2}$ from. Hence before attempting a length
based attack in this case the attacker has to compute the centraliser of $%
L_{B},R_{B}$.\medskip

So a third method to select the subgroups is

$A$ chooses $L_{A}=G,a_{1}\in G$, and publishes $L_{B}\subseteq
C_{G}(a_{1}),L_{B}=\{\alpha _{1},...,\alpha _{k}\}$,

$B$ chooses $L_{B}=G,b_{2}\in G$, and publishes $R_{A}\subseteq
C_{G}(b_{2}),R_{A}=\{\beta _{1},...,\beta _{k}\}$,

Hence $A$ chooses $(a_{1},a_{2})\in G\times C_{G}(b_{2})$ and publishes the
subgroup(s) as a set of generators of the centralisers $B$ chooses $%
(b_{1},b_{2})\in C_{G}(a_{1})\times G$, and hence can compute $x$ etc. Again
there is no explicit indication of where to select $a_{1}$ and/or $b_{2}$
from. Hence before attempting a length based attack in this case the
attacker has to compute the centraliser of $L_{B}$ and/or $R_{B}$.

a) If $z\neq e$ we require the following conditions 
\begin{equation}
\begin{array}[t]{ll}
\lbrack L_{A},L_{B}]=1, & [R_{A},R_{B}]=1, \\ 
\lbrack L_{B},Z]\neq 1, & [L_{A},Z]\neq 1, \\ 
\lbrack R_{B},Z]\neq 1, & [R_{A},Z]\neq 1, \\ 
\lbrack L_{A},R_{A}]\neq 1, & [L_{B},R_{B}]\neq 1.
\end{array}
\tag{2}
\end{equation}
All the above conditions for $z\neq e$ can arise by generalising from
properties of subgroups used in the SDG or CKLHC schemes for example the
second and third conditions in (2) arise from the observations that in
general $[LB_{n},B_{n}]\neq 1,[LB_{n},UB_{n}]=1$.\medskip \newline
b) If $z=e$ we require the following conditions 
\begin{equation}
\begin{array}[t]{ll}
\lbrack L_{A},L_{B}]=1, & [R_{A},R_{B}]=1, \\ 
\lbrack L_{A},R_{A}]\neq 1, & [L_{B},R_{B}]\neq 1, \\ 
\lbrack L_{B},R_{A}]\neq 1, & [L_{A},R_{B}]\neq 1.
\end{array}
\tag{3}
\end{equation}
.

\begin{itemize}
\item  Choose $z\in B_{n}$.\newline
ii) A(lice) chooses a secret braid $a_{1}\in L_{A}$, $a_{2}\in R_{A}$, her
private key; she publishes $K_{A}=a_{1}za_{2}$; the pair $(w,K_{A})$ is the
public key.\newline
i) B(ob) chooses a secret braid $b_{1}\in L_{B}$, $b_{2}\in R_{B}$, her
private key; she publishes $K_{B}=b_{1}zb_{2}$; the pair $(w,K_{B})$ is the
public key.\newline
iii) A and B can compute the common shared secret key $\kappa $ as $\kappa
=a_{1}K_{B}a_{2}$ and $\kappa =b_{1}K_{A}b_{2}$ respectively.\newline
i) Choose a public $w$
\end{itemize}

$h$ is a fixed collision-free hash function from braids to sequences of 0's
and 1's or, possibly, to braids, for which this choice for $h$. Again the
above protocol is considered with the commutativity conditions 2 or 3. \
Note the braids $K_{A}$ and $K_{B}$ are rewritten for example a normal form
to make the protocol secure. Full detail are given in \cite{Chowdhury}. It
was shown in \cite{Chowdhury} that the above key agreement protocol is a
generalisation of the key protocols given in \cite{Cha et al},\cite{Ko et al}%
,\cite{Shpilrain et al},\cite{Shpilrain2 et al}.\medskip

\underline{5. The Diffie-Hellman Decomposition Problem is Equivalent to the}

\underline{Multiple Simultaneous Conjugacy Search Problem}\medskip

In this section we will show that the DH-DP\ is equivalent to the MSCSP in
our key agreement protocol in section 2.1 hence showing the key exchange
protocols given in \cite{Cha et al},\cite{Ko et al},\cite{Stickel} are based
on the MSCSP and the key exchange protocols in \cite{Shpilrain et al},\cite
{Shpilrain2 et al} may be based on the MSCSP. In the braid group there are
various algorithms MSCSP can be solved with non-negligible probability such
as length based algorithms or the algorithm using ultra summit sets given in 
\cite{Gebhardt} so for braid group implementations the algorithms we show
are based on the MSCSP should not be used. Our result also applies to the
variant key exchange protocol and variant authentication scheme given in 
\cite{Chowdhury}.

We now introduce the concept of a CE (conjugacy extractor) function which we
build our attack upon.\medskip

Notation-We define a CE (conjugacy extractor) function to be a function that
on input of information from an user and information transmitted in a
cryptographic protocol gives as its output a conjugacy equation, by
conjugacy equation we mean an instantiation of the CSP. We denote $i$ CE
functions as CE$_{i}$, or CE if there is just one function
involved.\medskip\ \medskip

\underline{Theorem 1} \medskip

Solving the DH-DP is equivalent solving the MSCSP assuming $y_{a}$ and/or $%
y_{b}$ are invertible elements in the DH-DP.\medskip

\underline{Proof}\medskip

The proof is for when considering commutativity conditions 2 or 3 in the
generalised protocol above, when condition 3 are used then $x=e$. Firstly
define the CE for a protocol based on the DH-DP as follows 
\begin{eqnarray*}
CE_{1}(R_{I},y_{a})
&=&y_{a}R_{I}y_{a}^{-1}=axbR_{I}b^{-1}x^{-1}a^{-1}=axR_{I}x^{-1}a^{-1},R_{I}%
\in D \\
CE_{2}(S_{I},y_{b})
&=&y_{b}S_{I}y_{b}^{-1}=cxdS_{I}d^{-1}x^{-1}c^{-1}=cxS_{I}x^{-1}c^{-1},S_{I}%
\in B \\
CE_{3}(R_{I}^{\prime },y_{a}) &=&y_{a}^{-1}R_{I}^{\prime
}y_{a}=b^{-1}x^{-1}a^{-1}R_{I}^{\prime }axb=b^{-1}x^{-1}R_{I}^{\prime
}xb,R_{I}^{\prime }\in C \\
CE_{4}(S_{I}^{^{\prime }},y_{b}) &=&y_{b}^{-1}S_{I}^{\prime
}y_{b}=d^{-1}x^{-1}c^{-1}S_{I}^{\prime }cxd=d^{-1}x^{-1}S_{I}^{\prime
}xd,S_{I}^{^{\prime }}\in A
\end{eqnarray*}
$R_{I}$ is chosen from the subset that commutes with the subset that the
secret $a$ in $y_{a}$ is chosen from. $S_{I}$ is chosen from the subset that
commutes with the subset that the secret $b$ in $y_{b}$ is chosen from etc.
Since all the parameters are known to compute the $CE_{1}$,...,$CE_{4}$ are
easily computable. Note it is sufficient for one $CE$ to exist to prove the
theorem but we may want to compute more than one $CE$ because their
difficulty may vary, for example one of the $CE$ of the protocols \cite
{Shpilrain2 et al} can be used in a known length attack. Obviously $R_{I}$
(in general) does not commute with $x$ (similarly $R_{I}$ in general does
not commute with $a$ when conditions 3 are used) as this would mean an
attacker could easily recover the common secret key. So for $1\leq I\leq u$
this shows that the key agreement protocol in \cite{Ko et al} is based on
the MSCSP for the secret in either $ax$ or the MSCSP in the secret $cx$. So $%
a$ or $c$ can be found by right multiplying by $x^{-1}$ which is publicly
known. Hence the protocols in \cite{Cha et al}, \cite{Ko et al},\cite
{Shpilrain et al},\cite{Shpilrain2 et al},\cite{Stickel} are based on an
example of the MSCSP as 
\begin{eqnarray*}
&&((R_{1},R_{2},...,R_{u}),(CE_{1}(R_{1},y_{a}),CE_{1}(R_{2},y_{a}),...,CE_{1}(R_{u},y_{a})))%
\text{with solution }ax. \\
&&((S_{1},S_{2},...,S_{u}),(CE_{2}(S_{1},y_{b}),CE_{2}(S_{2},y_{b}),...,CE_{2}(S_{u},y_{b})))%
\text{with solution }cx. \\
&&((R_{1}^{\prime },R_{2}^{\prime },...,R_{u}^{\prime
}),(CE_{1}(R_{1}^{\prime },y_{a}),CE_{1}(R_{2}^{\prime
},y_{a}),...,CE_{1}(R_{u}^{\prime },y_{a})))\text{with solution }%
b^{-1}x^{-1}. \\
&&((S_{1}^{\prime },S_{2}^{\prime },...,S_{u}^{\prime
}),(CE_{2}(S_{1}^{\prime },y_{b}),CE_{2}(S_{2}^{\prime
},y_{b}),...,CE_{2}(S_{u}^{\prime },y_{b})))\text{with solution }%
d^{-1}x^{-1}.
\end{eqnarray*}
We now give applications of our theorem for the protocols \cite{Cha et al}, 
\cite{Ko et al},\cite{Shpilrain et al},\cite{Shpilrain2 et al},\cite{Stickel}
there are algorithm that solve the MSCSP with non-negligible probability
such as a length attack \cite{Garber et al}, so a length attack may be used
for the protocol \cite{Cha et al},\cite{Ko et al},\cite{Shpilrain et al}, 
\cite{Shpilrain2 et al}. Following the notation in \cite{Shpilrain2 et al} ,
where $G$ is a group so the security of the protocol in \cite{Shpilrain2 et
al} is always based on the MSCSP (because we know the generators for the
elements $a_{2}$ and $b_{1}$ we can use a length attack so disproving the
claim in \cite{Shpilrain2 et al} ), we have 
\begin{eqnarray*}
CE_{1}(A_{I},P_{A})
&=&P_{A}^{-1}A_{I}P_{A}=a_{2}^{-1}w^{-1}a_{1}^{-1}A_{I}a_{1}wa_{2}=a_{2}w^{-1}A_{I}wa_{2},
\\
A_{I} &\in &A\text{ , }A\text{ is a subgroup of }C_{G}(a_{1}) \\
CE_{2}(B_{I},P_{B})
&=&P_{B}B_{I}P_{B}^{-1}=b_{1}wb_{2}B_{I}b_{2}^{-1}w^{-1}b_{1}^{-1}=b_{1}wB_{I}w^{-1}b_{1}^{-1},
\\
R_{I} &\in &B\text{, }B\text{ is a subgroup of }C_{G}(b_{2})
\end{eqnarray*}
if it may be easy for a some sets $\{g_{1},g_{2},...,g_{k}\}$ of the
elements of $G$ to compute a part of or all (if $G$ is the braid group there
are algorithms that will compute a large part of the centraliser) 
\begin{equation*}
C(g_{1},...,g_{k})=C(g_{1})\cap ...C(g_{k})
\end{equation*}
then we can compute the following 
\begin{eqnarray*}
CE_{3}(C_{I},P_{A})
&=&P_{A}C_{I}P_{A}^{-1}=a_{1}wa_{2}C_{I}a_{2}^{-1}w^{-1}a_{1}^{-1}=a_{1}wA_{I}w^{-1}a_{1}^{-1},
\\
C_{I} &\in &C\text{,}C\text{ a subgroup of }C_{G}(b_{2}) \\
CE_{4}(D_{I},P_{B})
&=&P_{B}^{-1}D_{I}P_{B}=b_{2}^{-1}w^{-1}b_{1}^{-1}D_{I}b_{1}wb_{2}=b_{2}^{-1}w^{-1}B_{I}wb_{2},
\\
R_{I} &\in &D,D\text{ a subgroup of }C_{G}(a_{1})
\end{eqnarray*}
Following the notation in \cite{Shpilrain et al} let the elements
transmitted by Alice and Bob be invertible then 
\begin{eqnarray*}
CE_{1}(E_{I})
&=&b_{1}^{-1}w^{-1}a_{1}^{-1}E_{I}a_{1}wb_{1}=b_{1}w^{-1}E_{I}wb_{1},E_{I}%
\in B \\
CE_{2}(F_{I})
&=&b_{2}wa_{2}F_{I}a_{2}^{-1}w^{-1}b_{2}^{-1}=b_{1}wF_{I}w^{-1}b_{1}^{-1},B_{I}\in A
\end{eqnarray*}
and so we can use this equation in a length attack. It may be asked how much
better is a length attack using the conjugacy equations above for \cite
{Shpilrain2 et al},\cite{Shpilrain et al} (and related protocols) compared
to the known length attacks on \cite{Shpilrain2 et al},\cite{Shpilrain et al}%
, and if the above equations can be used to improve the known length attacks
(for example we may try using one and/or both equations above they can be
used to decide peeling off the correct generator in combination with the
algorithm that decides to peel of generators in an existing attack, an
example would be if the above existing algorithm is unable (i.e. pick at
random) to decide which is the correct generator to peel then peeling from $%
CE_{1}$ and /or $CE_{2}$ may be used to decide the correct generator).
Following the notation in\ \cite{Stickel} we have 
\begin{equation*}
CE_{1}(G_{I},c)=cG_{I}c^{-1}=a^{r}b^{s}G_{I}b^{-s}a^{-r}=a^{r}b^{\alpha
}a^{-r},G_{I}=b^{\alpha }\text{, for some }\alpha \text{ chosen by attacker}
\end{equation*}
or using the suggestion of using the element $e$ in \cite{Stickel} we have 
\begin{eqnarray*}
CE_{2}(H_{I},c)
&=&cG_{I}c^{-1}=a^{r}eb^{s}H_{I}b^{-s}e^{-1}a^{-r}=a^{r}(eb^{\alpha
}e^{-1})a^{-r},H_{I}=b^{\alpha }\text{,} \\
&&\text{for some }\alpha \text{ chosen by attacker.}
\end{eqnarray*}
Following the notation in \cite{Ko et al} we have 
\begin{eqnarray*}
CE_{1}(K_{I},y_{1})
&=&y_{1}K_{I}y_{1}^{-1}=axa^{-1}K_{I}ax^{-1}a^{-1}=axK_{I}x^{-1}a^{-1},K_{I}%
\in RB_{r} \\
CE_{2}(L_{I},y_{2})
&=&y_{2}L_{I}y_{2}^{-1}=bxb^{-1}L_{I}bx^{-1}b^{-1}=bxL_{I}x^{-1}b^{-1},L_{I}%
\in LB_{l} \\
CE_{3}(M_{I},y_{1})
&=&y_{1}^{-1}M_{I}y_{1}=ax^{-1}a^{-1}M_{I}axa^{-1}=ax^{-1}M_{I}xa^{-1},K_{I}%
\in RB_{r} \\
CE_{4}(N_{I},y_{2})
&=&y_{2}^{-1}L_{I}y_{2}=bx^{-1}b^{-1}N_{I}bxb^{-1}=bx^{-1}N_{I}xb^{-1},L_{I}%
\in LB_{l}
\end{eqnarray*}
Following the notation in \cite{THC} we have 
\begin{eqnarray*}
&&CE_{1}(T_{I},w_{i})=w_{i}T_{I}w_{i}^{-1}=y_{i-1}v_{i}y_{i}^{-1}K_{I}y_{i}v_{i}^{-1}y_{i-1}^{-1}=y_{i-1}v_{i}K_{I}v_{i}^{-1}y_{i-1}^{-1}
\\
&&CE_{1}(T_{I},w_{i}^{-1})=w_{i}^{-1}T_{I}w_{i}=y_{i}v_{i}^{-1}y_{i-1}^{-1}K_{I}y_{i-1}v_{i}y_{i}^{-1}=y_{i}v_{i}^{-1}K_{I}v_{i}y_{i}^{-1}
\\
&&K_{I}\text{ is chosen from the subgroup that generates the elements }x_{j}
\\
&&U_{I}\text{ is chosen from the subgroup that generates the elements }y_{J}
\\
&&\text{So the secrets }y_{1},...,y_{k}\text{ can be recovered and hence }%
v_{i}=y_{i-1}^{-1}w_{i}y_{i} \\
&&CE_{2}(U_{I},w)=wU_{I}w^{-1}=x_{0}v_{1}x_{1}v_{2}...v_{k}x_{k}U_{I}x_{k}^{-1}v_{k}^{-1}...v_{2}^{-1}x_{1}^{-1}v_{1}^{-1}x_{0}^{-1}
\\
&=&x_{0}v_{1}x_{1}v_{2}...v_{k}U_{I}v_{k}^{-1}...v_{2}^{-1}x_{1}^{-1}v_{1}^{-1}x_{0}^{-1}
\\
&&\text{recovering }x_{0}v_{1}x_{1}v_{2}...v_{k}\text{ gives }%
x_{k}=(x_{0}v_{1}x_{1}v_{2}...v_{k})^{-1}w\text{ similarly} \\
&&CE_{2}(U_{I},w^{-1})=w^{-1}U_{I}w\text{ gives }x_{0}\text{. Because all }%
v_{i}\text{ can be recovered} \\
&&\text{ so similarly repeating the attack above using } \\
&&CE_{2}(U_{I},(x_{0}v_{1})^{-1}wx_{k}^{-1}v_{k}^{-1})\text{ for }x_{1}\text{%
,}x_{k-1}\text{ and similarly by repeating} \\
&&\text{again all the }x_{i}\text{ can be recovered. }
\end{eqnarray*}
\bigskip Following the notation in \cite{Cha et al} we have 
\begin{eqnarray*}
CE_{1}(P_{I},c_{1})
&=&c_{1}P_{I}c_{1}^{-1}=a_{1}xa_{2}P_{I}a_{2}^{-1}x^{-1}a_{1}^{-1}=a_{1}xP_{I}x^{-1}a_{2}^{-1},P_{I}\in UB_{r}
\\
CE_{1}(Q_{I},y_{1})
&=&c_{2}Q_{I}c_{2}^{-1}=b_{1}xb_{2}Q_{I}b_{2}^{-1}x^{-1}b_{1}^{-1}=b_{1}xQ_{I}x^{-1}b_{1}^{-1},K_{I}\in LB_{r}
\end{eqnarray*}
The CKLHC protocol in \cite{Cha et al} was introduced as an improvement of
the KLCHKP protocol which it is a generalisation/modification of but we have
shown the CKLHC does not improve the security of the KLCHKP protocol as they
are both based on the MSCSP. This means using the KLCHKP and CKLHC
cryptosystems is no more secure than using the AAG (Anshel-Anshel-Goldfeld)
scheme \cite{AAG} in the connection that they can all be broken using by
solving the MSCSP. Hence this means there is no need to use the CKLHC
cryptosystem any longer. The theorem implies the Turing reduction of the
DH-DP to the MSCSP (MSCSP $\varpropto _{T}$ DH-DP) for the case when the
above DH-DP have related solutions, clearly a conjugacy extractor can be
feasibly computed- that is in polynomial time and polynomial space (for
parameters used in the CKLHC cryptosystem) using a finite number of group
operations. If the conjugacy extractor is not computable in polynomial time
and polynomial space in the connection of breaking a cryptographic protocol
then the above protocol may be secure from an attack by solving the MSCSP,
we may consider a generalisation of the MSCSP in the above cryptographic
protocol where $G$ is a semigroup instead of a group. Note if we have one $CE
$ then we exactly have the Turing reduction of the DP to the MSCSP (MSCSP $%
\varpropto _{T}$DP) , and hence the Turing reduction of the DH-DP to the
MSCSP (MSCSP $\varpropto _{T}$ DH-DP).

In \cite{Dehornoy} an authentication scheme is given based on the problem of
shifted conjugacy search problem (SCSP). It is not stated in \cite{Dehornoy}
not to select $r$ (the random value used in the commitment $x=r\ast p=r\cdot
dp\cdot \sigma _{1}\cdot dr^{-1}$) from a publicly known subgroup. Then an
attack is as follows.\medskip

1. Suppose $r\in R$ where $R=\{\alpha _{1},...,\alpha _{k}\}$ is a publicly
known subgroup of $B_{n}$. In this step it is required the attacker just
needs to find one element that commutes with $r$ and not al least with $%
dp\cdot \sigma _{1}$ (using a chosen algorithm by the attacker) to show that
SCSP can be reduced to solving the CSP. The attacker picks a subgroup of $R$
given by the generators $g_{1},...,g_{k}$. Then the attacker computes all of
or a large part of 
\begin{equation*}
N=C(\alpha _{1},...,\alpha _{k})=C(\alpha _{1})\cap ...C(\alpha _{k})\text{.}
\end{equation*}
2. Then 
\begin{eqnarray*}
&&CE_{1}(N_{I},r\ast p)=(r\cdot dp\cdot \sigma _{1}\cdot
dr^{-1})^{-1}N_{I}(r\cdot dp\cdot \sigma _{1}\cdot dr^{-1})= \\
&&dr\cdot \sigma _{1}^{-1}\cdot dp^{-1}\cdot r^{-1}\cdot N_{I}\cdot (r\cdot
dp\cdot \sigma _{1}\cdot dr^{-1})=dr\cdot \sigma _{1}^{-1}\cdot
dp^{-1}N_{I}dp\cdot \sigma _{1}\cdot dr^{-1}
\end{eqnarray*}
will be true.$\ N_{I}\in N$,$\ 1\leq I\leq M$. The the protocol can be based
on the MSCSP with 
\begin{eqnarray*}
&&((N_{1},N_{2},...,N_{M}),(CE(N_{1},r\ast p),CE(N_{2},r\ast
p),...,CE(N_{M},r\ast p)))\text{ } \\
&&\text{with solution }(dr\cdot \sigma _{1}^{-1}\cdot dp^{-1},dp\cdot \sigma
_{1}\cdot dr^{-1})\text{, }O=dp\cdot \sigma _{1}\cdot dr^{-1}.
\end{eqnarray*}
and $r$ can be found by computing $(\sigma _{1}^{-1}\cdot dp^{-1}O)^{-1}=r$,
there is a similar attack if $s$ (Alice's private key) is chosen from a
subgroup that is publicly known. Note the similar attack with $N_{I}$
commuting $r\cdot dp$ would mean the SCSP is just the CSP (no extra
computation using $d$ is required).

As a variant of the above algorithm an attacker may try to compute an
element $N_{I}^{^{\prime }}\in C(L)$ then it may be possible to use $%
N_{I}^{^{\prime }}$ instead of $N_{I}$ in the attack above where $L=r\ast p$
or $L=r\ast p^{^{\prime }}$, so in this variant knowledge of $s$ $%
(L=p^{^{\prime }}$ here) or $r$ being chosen from a subgroup is not
required. A different second CE on the authentication scheme in \cite
{Dehornoy} or the SCSP, is suppose in a general case we have a pair of
examples of the SCSP that have the same secret element $x=r\ast
p,x^{^{\prime }}=r\ast p^{^{\prime }}$ (the notation here follows \cite
{Dehornoy} with the secret element $r$) then 
\begin{eqnarray*}
CE_{2}(x,x^{^{\prime }-1}) &=& \\
CE_{2}(r\ast p,r\ast p^{^{\prime }}) &=&x\cdot x^{^{\prime }-1} \\
&=&(r\cdot dp\cdot \sigma _{1}\cdot dr^{-1})\cdot (r\cdot dp^{^{\prime
}}\cdot \sigma _{1}\cdot dr^{-1})^{-1} \\
&=&(r\cdot dp\cdot \sigma _{1}\cdot dr^{-1})\cdot dr\cdot \sigma
_{1}^{-1}\cdot dp^{^{\prime }-1}\cdot r^{-1} \\
&=&r\cdot dp\cdot dp^{^{\prime }-1}\cdot r^{-1}
\end{eqnarray*}
so the secret $r$ can be found by solving the CSP pair ($dp\cdot
dp^{^{\prime }-1},CE(r\ast p,r\ast p^{^{\prime }})$) for $r$, there is a
similar CE for $dr$ (use $x^{-1}\cdot x^{^{\prime }}$ instead of $x\cdot
x^{^{\prime }-1}$ and then $dr$ is transformed to $r$ etc). Then the attack
waits until $b=1$ is so in this case Alice send to Bob $r\ast s=r\cdot
ds\cdot \sigma _{1}\cdot dr^{-1}$ hence the attacker computes the private
key $s=r^{-1}\cdot (r\ast s)\cdot dr\cdot \sigma _{1}^{-1}$. We note the
above attacks is can be used to answer question 2.6 in \cite{Dehornoy} (with 
$y=p$ in th CSP). We note our attack can be used to solve the shifted
conjugacy decision problem. Our results suggest that CE functions should be
hard to compute hence semigroups may be considered because then the theorem
1 may be false because the elements $y_{a}$ and/or $y_{b}$ are not
invertible. This suggestion applies to any hard problem such as the EDL
problem below. Note in the algorithms in CE computations it may be
centraliser element(s) (call these $p_{i}$) that multiply the secret(s)
cancel out and it can be shown these factors in the centraliser are
efficiently computable, for example one way to do this is if $p_{i}$ is a
power of the fundamental braid then we can estimate a power of the
fundamental braid from the public elements (for example using a length
function) and so recover a $p_{i}$, or instead find this power by brute
force.

We now consider a problem related to the SCSP/CSP which is given a semigroup 
$G$, and publicly known functions $u:G\rightarrow G_{1},v:$ $%
G_{2}\rightarrow G,w:G_{3}\rightarrow G$, $G_{1},G_{2},G_{3}$ are subsets of 
$G$, and publicly the publicly known elements $y_{i}=u(r)v(p_{i})w(r^{-1}),$ 
$v(p_{i}),$ $1\leq i\leq n$ find if the element $r$. We observe that the
problem generalises the twisted conjugacy problem \cite{Staecker} and the
doubly twisted conjugacy problem \cite{Staecker}, e.g. with $i=1,u$ an
endomorphism, $v,w$ the identity map we recover the twisted conjugacy
problem, $G$ a group, $G_{1}=G,G_{2}=G,G_{3}=G$; so we refer to the above
problem as the GTCP (generalised twisted conjugacy problem) which we now
describe solutions for. Now consider the GTCP with $i>1$, select a pair $i,j$
with $v(p_{i})\neq v(p_{j})$,$1\leq i,j\leq n$ we have the conjugacy
extractor 
\begin{eqnarray*}
CE_{1}(y_{i},y_{j}) &=&y_{i}\cdot y_{j}^{-1} \\
CE_{1}(u(r)v(p_{i})w(r^{-1}),u(r)v(p_{j})w(r^{-1}))
&=&u(r)v(p_{i})w(r^{-1})\cdot (u(r)v(p_{j})w(r^{-1}))^{-1} \\
&=&u(r)v(p_{i})w(r^{-1})\cdot w(r^{-1})^{-1}v(p_{j})^{-1}u(r)^{-1} \\
&=&u(r)v(p_{i})v(p_{j})^{-1}u(r)^{-1}
\end{eqnarray*}
so here we solve the CSP pair $(v(p_{i})v(p_{j})^{-1}$,$CE_{1}(y_{i},y_{j})).
$ Once $u(r)$ is obtained (attempt to) use the inverse of $u$ to get $r$. 
\begin{eqnarray*}
CE_{2}(y_{i},y_{j}) &=&y_{j}^{-1}\cdot y_{i} \\
CE_{2}(u(r)v(p_{i})w(r^{-1}),u(r)v(p_{j})w(r^{-1}))
&=&(u(r)v(p_{j})w(r^{-1}))^{-1}\cdot u(r)v(p_{i})w(r^{-1}) \\
&=&w(r^{-1})^{-1}v(p_{j})^{-1}u(r)^{-1}\cdot u(r)v(p_{i})w(r^{-1}) \\
&=&w(r^{-1})^{-1}v(p_{j})^{-1}v(p_{i})w(r^{-1})
\end{eqnarray*}
so here we solve the CSP pair $(v(p_{j})^{-1}v(p_{i})$,$CE_{2}(y_{i},y_{j})).
$ Once $w(r)$ is obtained use the inverse of $w$ to get $r$. Note for one of
the above $CE$ functions for the twisted conjugacy problem the problem is
just the conjugacy search problem. The above can be repeated for different $%
i,j$ to get a reduction to the MSCSP. Another method to solve the GTCP as
follows.\smallskip 

1. Suppose $G_{1}=\{\alpha _{1},...,\alpha _{k}\},G_{3}=\{\beta
_{1},...,\beta _{l}\}$ are publicly known. In this step it is required that
one element that commutes with $r$ is to be found. Pick subgroups of $%
G_{1},G_{3}$ given by the generators $g_{1},...,g_{L},h_{1},...,h_{L}.$ Then
compute all of or a large part of 
\begin{eqnarray*}
N &=&C(\alpha _{1},...,\alpha _{k})=C(\alpha _{1})\cap ...C(\alpha _{k})%
\text{.} \\
M &=&C(\beta _{1},...,\beta _{k})=C(\beta _{1})\cap ...C(\beta _{l})\text{.}
\end{eqnarray*}
2. Then 
\begin{eqnarray*}
&&CE_{3}(M_{I},y_{i})=u(r)v(p_{i})w(r^{-1})M_{I}(u(r)v(p_{i})w(r^{-1}))^{-1}=
\\
&&u(r)v(p_{i})w(r^{-1})M_{I}w(r^{-1})^{-1}v(p_{i})^{-1}u(r)^{-1}= \\
&&u(r)v(p_{i})M_{I}v(p_{i})^{-1}u(r)^{-1}
\end{eqnarray*}
and 
\begin{eqnarray*}
&&CE_{4}(N_{I},y_{i})=(u(r)v(p_{i})w(r^{-1}))^{-1}N_{J}u(r)v(p_{i})w(r^{-1})=
\\
&&w(r^{-1})^{-1}v(p_{j})^{-1}N_{J}v(p_{i})w(r^{-1}))
\end{eqnarray*}
$N_{J}\in N,M_{I}\in M,\ 1\leq I\leq m,1\leq J\leq n$. Hence using $%
CE_{3}(M_{I},y_{i}),CE_{4}(N_{I},y_{i})$: the GTCP has a solution
respectively in the MSCSPs with 
\begin{eqnarray*}
&&((N_{1},N_{2},...,N_{n}),(CE_{4}(N_{1},y_{j_{1}}),CE_{4}(N_{2},y_{j_{2}}),...,CE_{4}(N_{n},y_{j_{n}})))%
\text{ } \\
&&\text{with solution }w(r^{-1})^{-1}v(p_{j})^{-1}\text{;}
\end{eqnarray*}
\begin{eqnarray*}
&&((M_{1},M_{2},...,M_{m}),(CE_{3}(M_{1},y_{i_{1}}),CE_{3}(M_{3},y_{i_{2}}),...,CE_{3}(M_{m},y_{i_{m}})))%
\text{ } \\
&&\text{with solution }u(r)v(p_{i})\text{;}
\end{eqnarray*}
; from $w(r^{-1})^{-1}v(p_{j})^{-1}$, $u(r)v(p_{i})$ we can obtain $r$ by
using a right multiplication and using the inverses of $w,u$; this show the
twisted conjugacy problem can be deterministically reduced to the MSCSP. We
observe the algorithm in section 7 below can be used to attempt to solve the
twisted or doubly twisted conjugacy problem with or without using $u$ or $v$%
. Observe once we have found a solutions to the twisted conjugacy problem,
SCSP this means we can solve the decision version of the twisted conjugacy
problem and SCSP. \smallskip 

\underline{6. A Solution for the EDL type problem in Non-commutative
Semigroups.}\medskip

The EDL braid type problem was proposed in \cite{Thomas} where it is assumed
to be hard. Following the notation in our definition above of the EDL we
have the following theorem.\medskip

\underline{Theorem 2}\medskip

Given (the DP equations) $y_{a}=uav$, $y_{b}=wbx$ in the EDL is sometimes
equivalent to solving the CSP if $y_{a}$,$y_{b}$ are both invertible
elements.\medskip

\underline{Proof}\medskip

We will solve this problem by solving a system of DP equations for certain
values of the secrets so our proof can be used to solve a system of DP
equations for example in showing above the shifted conjugacy based protocol
is based on the CSP. Assume $w=u$ and $x=v$,$a\neq b$, then $y_{a}=uav$, $%
y_{b}=ubv$, and so the $CE$ functions give 
\begin{eqnarray*}
y_{a}y_{b}^{-1} &=&uab^{-1}u^{-1} \\
y_{a}^{-1}y_{b} &=&v^{-1}a^{-1}bv
\end{eqnarray*}
and so we can solve the CSP pairs for $(ab^{-1},y_{a}y_{b}^{-1})$, and $%
(a^{-1}b,y_{a}^{-1}y_{b})$ for $g_{1}=u$ and $g_{2}=v$ respectively using an
algorithm for the CSP. The the verification (can be done efficiently using
the algorithms for the word problem when $G$ is the braid group e.g. see 
\cite{Cha et al}) $y_{a}=g_{1}ag_{2}$, $y_{b}=g_{1}bg_{2}$ will be true by
the above assumption, hence we have shown the EDL to be true in this case.
For some examples of the CSP there are fast algorithm for it for example see 
\cite{Gebhardt} hence the assumption in \cite{Thomas} that the EDL is hard
is not always true. If we know the generators of the subgroups A and B then
we may use a length based algorithm to recover the secret element with
non-negligible probability.

We re-define again the EDL problem more generally as follows

\textit{Public information}: $G$ is a Semigroup.$A,B,C,D$ are subsets of $G.$
$a_{i},b_{i},x_{i},y_{i}\in G$ with $y_{i}=a_{i}x_{i}b_{i}$, $1\leq i\leq m$%
\newline
\textit{Secret Information}: $a_{i}\in A_{i}$, $b_{i}\in B_{i}$, ($A_{i}$
and $B_{i}$ are subgroups).\newline
\textit{Objective}: Decide if $F_{x_{1}}(y_{_{1}})\cap
F_{x_{2}}(y_{_{2}})...\cap F_{x_{r}}(y_{_{r}})\neq \varnothing $.. Where $%
F_{\beta }(\alpha )=\{(a,b)\in B_{n}\times B_{n}$ :$\alpha =a\beta b\}.$%
\medskip

\underline{Theorem 3}\medskip

Given the generalised EDL above is sometimes equivalent to solving the CSP.
The generalised EDL may be partially solved in the connection a subset
integers $t_{1},t_{2},...,t_{r}$ in $[1,m]$ we can decide if $%
F_{x_{t_{1}}}(y_{_{t_{1}}})\cap F_{x_{t_{2}}}(y_{_{t_{2}}})...\cap
F_{x_{t_{r}}}(y_{_{t_{r}}})\neq \varnothing $.\medskip

\underline{Proof}\medskip

Again we will solve this problem by solving the a system of DP equations for
certain values of the secrets so our proof can be used to solve a system of
DP equations. Assume $a_{i}=a_{j}$ and $b_{i}=b_{j}$ for all $i,j\in \lbrack
1,m]$ and $i\neq j$ then $y_{i}=a_{i}x_{i}b_{i}$, $y_{j}=a_{j}x_{j}b_{j}$,
and so the CE functions give 
\begin{eqnarray*}
y_{i}y_{j}^{-1} &=&a_{i}x_{i}x_{j}^{-1}a_{i}^{-1} \\
y_{i}^{-1}y_{j} &=&b_{i}^{-1}x_{i}^{-1}x_{j}b_{i}
\end{eqnarray*}
and so we can solve the CSP pairs $(x_{i}x_{j}^{-1},y_{i}y_{j}^{-1})$, and $%
(x_{i}^{-1}x_{j},y_{i}^{-1}y_{j})$ for the solutions $g_{1}=a_{i}$ and $%
g_{2}=b_{i}$ respectively using an algorithm for the CSP. We can get more
conjugacy extractor functions by choosing different values for $i$ and $j$.
The the verification (can be done efficiently using the algorithms for the
word problem when $G$ is the braid group e.g. see \cite{Cha et al}) $%
y_{a}=^{?}g_{1}x_{a}g_{2}$, $y_{b}=^{?}g_{1}x_{b}g_{2}$ will be true by the
above assumption for all $(a,b)=(i,j)$, hence we have shown the EDL to be
true in this case.

If we know the generators of the subgroups $A$ and $B$ then we may use a
length based algorithm to recover the secret element with non-negligible
probability.

The EDL can be partially solved if it is true that the assumption $%
a_{i}=a_{j}$ and $b_{i}=b_{j}$ for at least two integers $i$ and $j$, $%
i,j\in \lbrack 1,m]$ and $i\neq j.$\ Then the proof that the EDL can be
partially solved is the same as above except there are fewer choices for $i$
and $j$.\medskip

\underline{7. Second Algorithm using CE Functions}\medskip

Given (the DP equation) $u=xaz$. This attack reveals partial information
about the secret $z$ or totally recover $z$. This attack is a generalisation
of our attack on the DP by using a MSCSP.

1. The attacker picks elements $S_{I}$ according to some criteria relating
to commutativity, for example elements $S_{I}$ may be picked randomly or $%
S_{I}$ may be composed of a few Artin generators as these may commute to
some degree with $z$.

2. Then for $1\leq I\leq M$ for a sequence of integers $T_{I}$%
\begin{equation*}
CE_{I}(S_{I},u)=uS_{I}u^{-1}=xazS_{I}z^{-1}a^{-1}x^{-1}=xa\overline{z}S_{I}%
\overline{z}^{-1}a^{-1}x^{-1}
\end{equation*}
where (with probability $\rho )$ $\overline{z}$ is a partial factor of $z$
for some $I$ this means $z=z_{T_{I}}\overline{z}_{T_{I}}$.

3. We solve for each $I$ the CSP $(S_{I},CE_{I}(S_{I},u))$ for the solution $%
xa\overline{z})$ and hence compute $z_{T_{I}}=((xa\overline{z}%
^{-1})^{-1}xaz)^{-1}$. Note if $S_{I}$ is selected from the centraliser of $%
z $ then we can use the MSCSP at this step (so this shows DP is Turing
reducible to MSCSP).

4. We now find (in some way) $z$ using the information $(S_{I},xa\overline{z}%
,z_{T_{I}})$ and the other information used in the protocol. One of the
simplest choices to implement this step is trying to find $\overline{z}$ for
each $I$ by brute force and hence possibly recover $z$.\medskip

A variant of the above attack is after $z_{T_{I}}$ is recovered is to repeat
at the attack (at least once) by iterating with $uz_{T_{I}}^{-1}$ instead of 
$u$ (and obviously all other values may be different) so in this way we may
be able to find a bigger factor of $z$. It may be true (with some
probability $\rho _{2}$) that $\overline{z}$ contains a partial factor of $a$
which means the CSP is solved to give $\overline{z}_{T_{I}}a_{T_{I}}$ where $%
a_{T_{I}}$ is some partial factor of $a$. Then the simplest choice at this
step to recover $z$ is to find $a_{T_{I}}$ by brute force and use $a_{T_{I}}$
to recover $z.\medskip $

\underline{Conclusion}\medskip

We have shown the protocols \cite{Cha et al},\cite{Ko et al},\cite{Shpilrain
et al},\cite{Shpilrain2 et al},\cite{Stickel},\cite{THC},\cite{Dehornoy} can
have security based on the MSCSP. We have shown the DP and DH-DP can be
solved by the MSCSP. Our theorem 1 implies that the CKLHC cryptosystem and
related cryptosystems are MSCSP based so are no more secure than using the
AAG protocol \cite{AAG}. Our theorem 1 implies that semigroups should (for $%
G $) be used for the protocol in \cite{Chowdhury} to be secure so not based
on the MSCSP. We should not use the CKLHC\ protocol in \cite{Cha et al} or
related protocols (which are suggested to be used in braid groups) compared
to using the AAG protocol as it is no more secure than using the AAG in the
connection they are based on the MSCSP.

\end{document}